\documentclass{elsart}
\usepackage{youngtab}
\usepackage{epsfig}
\def\by{\begin{Young}}
\def\ey{\end{Young}}
\def\be{\begin{equation}}
\def\ee{\end{equation}}
\def\bea{\begin{eqnarray}}
\def\eea{\end{eqnarray}}
\def\beq{\begin{equation}}
\def\eeq{\end{equation}}
\def\beqa{\begin{eqnarray}}
\def\eeqa{\end{eqnarray}}
\def\bce{\begin{center}}
\def\ece{\end{center}}

\def\P{{\sf P}}

\def\<{\langle}
\def\>{\rangle}

\def\half{{\textstyle{1\over 2}}}
\def\ohalf{{\textstyle{1\over 2}}}
\def\thalf{{\textstyle{3\over 2}}}

\def\thalf{{\textstyle{3\over 2}}}

\newcommand{\bra}[1]{\langle {#1} |}                        
\newcommand{\ket}[1]{| {#1} \rangle}                        
\def\vqhalf{{\textstyle{\vec{Q}\over 2}}}
\def\qhalf{{\textstyle{Q\over 2}}}

\newcommand{\AmS}{{\protect\the\textfont2
  A\kern-.1667em\lower.5ex\hbox{M}\kern-.125emS}}
\Yvcentermath{1}

\begin{document}
\begin{frontmatter}
\title{Baryon Magnetic Moments in Relativistic Quark Models}

\author{B. Juli\'a-D\'{\i}az}
\and
\author{D. O. Riska}

\address{Helsinki Institute of Physics and
	  Department of Physical Sciences, \\
        POB 64, 00014 University of Helsinki, Finland}

\begin{abstract}
It is shown that the phenomenological description
of the baryon magnetic moments in the quark model 
carries over to the Poincar\'e covariant extension of the model.
This applies to all the three common forms of 
relativistic kinematics with
structureless constituent currents, which
are covariant under the corresponding kinematic subgroups.
In instant and front form kinematics the calculated magnetic
moments depend strongly on the 
constituent masses, while in point form
kinematics the magnetic moments are fairly
insensitive to both the quark masses and the wave function model.
The baryon charge radii and 
magnetic moments are determined in the different forms of
kinematics for the light-flavor, strange and charm
hyperons. The wave function model
is determined by a fit to the electromagnetic form
factor of the proton.   
\end{abstract}
\end{frontmatter}

\section{Introduction}

The prominence of the non-relativistic
quark model is partly due to the simple and qualitatively
satisfactory description of the 
masses and the magnetic moments of the ground state
baryons, which it provides. Notwithstanding this satisfactory
phenomenology the model 
lacks dynamical consistency, as the high velocity of 
the confined quarks a priori invalidates the non-relativistic 
description of the constituent quarks. 
This problem may be overcome by recasting the model 
into Poincar\'e covariant form. It is shown here that the
satisfactory description of the baryon magnetic
moments carries over to this relativistic
version of the quark model at the modest price
of somewhat smaller constituent masses than commonly used,
provided that the baryon wave function is determined by a
fit to the electromagnetic form factors of the nucleon.

The calculation of the electromagnetic observables of baryons
in the relativistic quark model calls for a choice of kinematic 
subgroup of the group of Poincar\'e transformations, a choice
which is referred to as the ``form of kinematics'', in addition 
to the choice of wave function and current operator models. 
The three forms of kinematics, originally outlined by Dirac~\cite{Dirac}, 
are commonly referred to as ``instant'', ``point'', and ``front'' 
form kinematics. The corresponding kinematic subgroups of the group 
of Poincar\'e transformations are the group of rotations and 
translations at fixed time $E(3)$, the group of Lorentz 
transformations $SO(1,3)$, and the symmetry group on the light cone. 
The present relativistic quantum mechanical
approach should be distinguished from the approach based
on relativistic field theory, in which classes of Feynman
diagrams are summed in Bethe-Salpeter type equations,
as e.g. in Ref~\cite{merten}. 

It was recently noted that
a fair description of the electromagnetic form 
factors of the nucleons 
over the empirically known range of momentum transfer
may be achieved
with all these three forms of relativistic kinematics
with simple algebraic wave function models for the confined 
three-quark system~\cite{Bruno}. The model employed 
structureless constituent current operators, that are 
covariant under transformations of the corresponding 
kinematic subgroup. That work was stimulated by the
fairly satisfactory results obtained in point form
kinematics for the electromagnetic form factors of the
nucleons with a very compact wave function model in
Ref.~\cite{plessas}.

Here a calculation of the baryon magnetic moments
and charge radii is carried out with the relativistic
constituent quark model in all the three forms of 
relativistic kinematics. Both the strange and the charm 
as well as the recently
discovered doubly charm hyperons are considered. 
The functional form of the spatial wave function model is 
taken to be similar for all 
three forms of kinematics, although the spatial extent 
of the wave function, which is required for a satisfactory 
description of the electromagnetic form factor of the proton, 
depends on the form of kinematics. 

With instant and front form kinematics 
the calculated baryon magnetic moments are found to
be very sensitive to both the
wave function model and the value of the constituent
mass. With point form kinematics the calculated magnetic
moments of the nucleon are on the other hand found to be quite insensitive to 
the constituent mass. 
Instead there is a sensitivity to the mass of the baryon, which
leads to smaller values for the calculated magnetic moments 
of the heavy hyperons than with the other forms of
kinematics. In instant and front form kinematics both
the empirical 
magnetic moments and the
charge radii of the strange hyperons are well reproduced with the
same spatial wave function, as used for the nucleons. In contrast,
reproduction of the empirical charge radius of the $\Sigma^-$
hyperon would in point form demand that its matter radius be
more than twice that of the nucleon. Even with such an
extended wave function model the calculated 
magnetic moments of most of the strange hyperons
remain somewhat small in point form kinematics.
 
In \cite{Bruno} it was noted that 
the relativistic quark model gives reasonable values 
for the nucleon even for zero constituent mass. By
considering the hyperon magnetic moments it is shown
here that, while the constituent masses of 
the light flavor quark masses may be taken to be very small,
agreement with the experimental magnetic moments requires
finite values for the strange quark mass.

This paper is organized in the 
following way: Section 2 contains a description
of the calculation of the magnetic moments in the Poincar\'e 
covariant quark model in the three different forms of relativistic 
kinematics. The wave function model is described in Section 3. 
In Section 4 results for the magnetic moments 
and the charge radii of the strange hyperons 
are given. The corresponding results for the 
charm hyperons are given in Section 5. 
A concluding discussion of the results is given in Section 6.

\section{Magnetic Moments and Charge Radii}  

\subsection{Definition of the magnetic moment}

The magnetic moment of a baryon may be
defined as the value of the invariant
magnetic form factor $G_M$ at zero invariant momentum transfer. 
The form factor is defined as a matrix element of an appropriate 
component of the electromagnetic current operator, which depends 
on the form of kinematics. In instant and point form, the magnetic 
moment of a spin $1/2$ baryon may thus be defined as the
matrix element of the spin raising current 
component $I_+ = (1/2)(I_x + i I_y)$ in 
the Breit frame:
\begin{equation}
\mu = G_M(0) = {\sqrt{1+\eta}\over \sqrt{\eta}}
\langle \half , \vqhalf \vert
I_+ (0) \vert -\half, -\vqhalf \rangle \, .
\label{GM}
\end{equation}
Here $\vec Q$ has been taken to be parallel to the $z-$axis 
and $\eta$ is defined as $\eta = Q^2/ 4 M^2$, where $M$ is the baryon mass. 
In the case of front form kinematics the appropriate definition 
is on the other hand:
\begin{equation}
\mu = G_M(0) = F_1(0) + F_2(0)\, ,
\label{GMF}
\end{equation}
where $F_1$ and $F_2$ are the Dirac and Pauli form factors of the
nucleon. These are in turn defined as the following matrix elements 
of the ``plus'' component
of the current $I^+ = I^0 + I^z$:
\begin{equation}
F_1(0) = \langle \half \,\vert I^+ \vert \, \half \rangle\, ,
\quad F_2(0)= {1\over \sqrt{\eta}}\langle{-\half}\,
\vert I^+ \vert {\half}\rangle \, .
\label{F1F2}
\end{equation}
In this case the momentum transfer has to be transverse to the
$z-$direction: $\vec Q = \vec Q_\perp$. Without loss of generality
$Q$ may be taken to lie in the direction of the $x-$axis
\cite{chung}. 

For spin $3/2$ baryons, as the $\Delta(1232)$ and the $\Omega^-$, 
the corresponding definition of the magnetic moment in 
instant and point form kinematics is:
\begin{equation}
\mu = G_M(0) = {\sqrt{3(1+\eta)}\over \sqrt{\eta}}
\langle \thalf ,\, \qhalf \vert
I_+ (0) \vert \half ,\, -\qhalf\rangle \, .
\label{GM32}
\end{equation}
In front form kinematics the definition is again:
\beq
\mu = G_M(0) = F_1(0) + F_2(0) \,,
\label{GM32F}
\eeq
but now $F_1$ and $F_2$ are obtained as \cite{schlumpf}:
\beqa
 F_1 &=& \bra{\thalf,P'} I^+(0) \ket{P,\thalf}\, , \nonumber \\
F_2  - 2 F_1  &=& \sqrt{3\over \eta} \bra{\ohalf,P'} I^+(0) \ket{P,\thalf} \,.
\label{F1F22}
\eeqa
Here $P$ and $P'$ are the initial and final total momenta.

\subsection{Definition of the charge radii}

The mean square charge radius of a baryon is defined as the
derivative of the charge (electric) form factor $G_E(Q^2)$ with
respect to the invariant squared momentum transfer as:
\beq
<r^2>  = -6 \bigg( {\d G_E (Q^2) \over \d Q^2} \bigg)_{Q^2=0}.
\eeq 
For charged baryons the convention is to divide out the
charge from the electric form factor in the definition of
the charge radius.
In the case of instant and point form kinematics
$G_E$ is defined as the matrix element:
\beq
G_E(Q^2) = \langle \sigma, \qhalf \vert
I_z (0) \vert \sigma, -\qhalf \rangle \, , 
\qquad \sigma=\half,\thalf\, .
\label{GE}
\eeq
In the case of front form kinematics the appropriate 
definition is \cite{chung}:
\beq
G_E(Q^2) = F_1(Q^2) - \eta F_2(Q^2)\, . 
\eeq
Here $F_1$ and $F_2$ are the Dirac form factors  
defined in Eq.~(\ref{F1F2}) for spin 1/2 baryons and 
in Eq.~(\ref{F1F22}) for spin 3/2 baryons.

\section{The wave function model in the magnetic form factors}

\subsection{The wave function model}

The baryon wave function models are here taken 
as products of a completely symmetric spatial component 
and $SU(N_f)_F$ symmetric spin-flavor state vectors
in the rest frame. 
The constituent quark masses appear in the current operators and
in the boosts as well as in the Wigner and Melosh
rotations, which connect the initial and final states to 
the spins and momenta that appear in the rest frame baryon wave 
functions. The assumption of a completely symmetric spatial
wave function for all systems of three quarks implies 
that the mass operator, the eigenfunctions of which are the 
wave functions, is symmetric in the quark masses as
well. Examples of mass operators of this
general form, with spectra that agree with the empirical
baryon spectra, are given in Ref.~\cite{codann}.

For the ground state baryons the spatial wave function will be
taken to be a completely symmetric $S-$state. 
This function will be taken to have the form:
\begin{equation}
\varphi_{0}(\P)= {\mathcal N} 
\left( 1+ {\P^2\over 4 b^2}\right)^{-a}\, ,
\label{ground}
\end{equation}
where $\P$ is the hyperspherical momentum, 
$\P:= \sqrt{2 (\vec\kappa^2+\vec q^2)}$, where
$\kappa$ and $q$ are the Jacobi coordinates for the
three-quark system. In (\ref{ground}) ${\mathcal N}$ is a normalization 
constant, while the exponent $a$ and
$b$ are adjustable parameters. The values of the parameters 
for the three different forms of relativistic kinematics
are chosen as in Ref.~\cite{Bruno}, 
and are listed in Table~\ref{parameters}.

\begin{table}[b]
\caption{The parameter values in the ground 
state wave function (\ref{ground}) used for the three different 
forms of kinematics. The corresponding matter radii
$r_0$ are listed in the last column.\label{parameters}}
\bce
\begin{tabular}{lcccc}
       & $m_q$ (MeV)  & $b$ (MeV) & $a$ &$r_0$ (fm) \\ 
\hline
instant form & 140       &  600        &   6  & 0.63  \\
point form   &  350           &  640        &  9/4 & 0.19 \\
front form   &  250           &  500        &   4  & 0.55 \\ 
\hline
\end{tabular}
\ece
\end{table}

The $SU(N_f)_F$ symmetry of the spin-flavor wave state
of the baryons in the rest frame is broken by the boosts to the
Breit frame, which change the spin quantization axis.
Both the boosts and the current operators in addition break
the flavor symmetry explicitly by the dependence on the 
constituent quark mass, which is different for quarks of different
flavor. 

The spin-flavor wave functions of 
the octet and decuplet baryons can be formally
expressed in terms of the flavor and spin symmetry
assignments as:
\beq
\varphi_{SF}(\ohalf)=
{\young(123)}_{\,SF}
=
{1\over \sqrt{2}} \Bigg\{ \, \vspace{-3mm}
{\young(12,3)} _{\,S}
{\young(12,3)} _{\,F}
+
{\young(13,2)} _{S\,}
{\young(13,2)} _{F\,}
\Bigg\} \, ,
\eeq
and 
\beq
\varphi_{SF}(\thalf)=
{\young(123)}_{\,SF}
=
{\young(123)} _{\,S}
{\young(123)} _{\,F} \, ,
\eeq
respectively. In the following we will make use of the
following notation:
\beq
\ket{S_{MS}}= {\young(12,3)} _S \, ,
\;
\ket{S_{MA}}= {\young(13,2)} _S\, ,
\;
\ket{S_S}= {\young(123)} _{\,S}\, ,
\; 
\ket{S_A}= {\young(1,2,3)} _{\,S}\, .
\eeq
In Table~\ref{bwf4} we give the explicit expressions for the 
mixed symmetry wave functions of the octet baryons.

\subsection{Magnetic form factors in the different forms
of kinematics}

In instant and in point form kinematics, the magnetic
form factors may be written as an integral over
the spectator momenta in the Breit frame as \cite{Bruno}:

\beq
G_M(\eta) = \int \d^3 p_2 \d^3 p_3\,  
\varphi\left({{\kappa'}^2+q^{'2}\over 2 b^2}\right)
\varphi\left({\kappa^2+q^2\over 2 b^2}\right)
{\mathcal I}_+(\vec p_2,\vec p_3)\; .
\label{GEM}
\eeq
Here the wave functions $\varphi$ have
the form (\ref{ground}),
and depend on the Jacobi momenta in the initial and final
rest frames respectively, and
${\mathcal I}_+$ is the matrix element of the spin
raising component
of the Wigner rotated Dirac current operator,
multiplied by a Jacobian factor that takes into account the
transition from the rest frame to the Breit frame. The
explicit expressions for the Wigner rotated 
current matrix element and the Jacobian factor that
applies in instant and
point form kinematics are given in Ref.~\cite{Bruno}.   
The Breit frame momenta $\vec p_i$ are related to the
initial and final state momenta by Lorentz boosts,
which depend on the form of kinematics.

In front form kinematics the magnetic form factors
are obtained as a linear combination of the
Dirac form factors of the baryons (\ref{GM32F}). The
expressions for the latter take the form \cite{Bruno,chung}:
\beqa
F_\alpha(Q^2) &=& \int _{0}^{1} \d\xi_1 \,\d\xi_2 \,\d\xi_3 \, 
{\delta(\xi_1+\xi_2+\xi_3-1)
 \over\xi_1\xi_2\xi_3}\,\nonumber \\
&& \int \d\vec{k}_{1\perp}\, \d\vec{k}_{2\perp}\, \d\vec{k}_{3\perp}\, 
\delta(\vec{k}_{1\perp}+\vec{k}_{2\perp}+\vec{k}_{3\perp})
{\omega_1 \omega_2 \omega_3 \over M_0}\, \nonumber\\
&&
\varphi^*(\xi_1,\vec{k}_{1\perp}',\xi_2,\vec{k}_{2\perp}',\xi_3,
\vec{k}_{3\perp}')
{\mathcal F}_\alpha(Q^2)
\varphi(\xi_1,\vec{k}_{1\perp},\xi_2,
\vec{k}_{2\perp},\xi_3,\vec{k}_{3\perp})\; .
\eeqa
Here the relations between the momentum variables
of the struck (``1'') and
the two spectator quarks (``2,3'') in the initial and
final states are:
\be
k_{1\perp}'= k_{1\perp}+(1-\xi_1)Q_\perp\; , \qquad 
k_{i\perp}'= k_{i\perp}-\xi_1Q_\perp \qquad i=2,3\; .
\ee
The variables $\xi_i$ are defined as $\xi_i = p_i^+/P^+$,
where $P$ is the total momentum and $+$ refers to the 
``plus'' component in the light cone representation.

The factors ${\mathcal F}_\alpha(Q^2)$ involve the spin-isospin amplitudes
and  the effects of the Melosh 
rotations on both spectators and the Dirac
current $I_x$ of the struck constituent. The explicit
expressions for these are given in Ref.~\cite{Bruno}. The factor
$\omega_1\omega_2\omega_3/M_0$  is the Jacobian factor for
front form kinematics. Here $\omega_i$ are single constituent
energies and $M_0$ is the free mass operator.

\section{Magnetic moments and charge radii of nucleons and strange hyperons}

There are relations between 
the magnetic moments of the baryons, that arise
from the spin-flavor matrix elements, which do not depend
on the form of relativistic kinematics. These relations involve 
matrix elements of 
the spin part of the current density operator, which depends 
on the specific form of kinematics through the boosts of the quark
momenta and the Wigner rotations on the quark spins. 
The spin-flavor matrix elements that enter in the calculation 
of the magnetic moments of the baryons may formally
be denoted as
$\bra{B} \; {\mathcal S F}  \; \ket{B}$. Here $\mathcal S$ is the spin operator,
which is determined by the Dirac current of the struck constituent
and the boosts (in the case of the transition moments) and the
Wigner (or Melosh in the case of front form
kinematics) rotations. The factor 
$\mathcal F$ is the flavor dependent charge operator of the struck quark:
\beq
 {\mathcal F} = \bigg\{ {\lambda_3 \over 2} + 
{\lambda_8\over 2\sqrt{3}} \bigg\} \, .
\label{charge}
\eeq
The spin operator $\mathcal S$ depends on flavor through the flavor
dependence of the quark mass.
Appendix~\ref{ap:me} contains the explicit expressions for the  
spin-flavor matrix elements that appear in the calculation of both 
the magnetic form factors and the electric charge radii once
the flavor part has been calculated using $SU(4)$ wave functions.
These expressions reduce to the well-known static
quark model expressions (e.g. \cite{Dannbom}), if the 
spin operator is assumed to have the static
form
\beq
{\mathcal S}_{\rm static}= \sum_i {q_i\over 2 m_i} \sigma_{iz} \,.
\label{nonrel}
\eeq

The expressions in Appendix~\ref{ap:me} contain matrix elements
of the spin part of the baryon wave functions. These matrix 
elements depend on the current density operator and on the 
form of relativistic kinematics.

In Table~\ref{tab:1} we give results for the magnetic moments 
that are obtained with equal flavor independent masses
for the constituent $u$, $d$ and $s$ quarks. For the 
three different forms of kinematics two cases are considered: 
one with the quark mass equal to zero, and a second one 
with non-zero quark mass. The relativistic quark
model differs from the non-relativistic quark model in
that finite values for the magnetic moments are obtained
even in the case of vanishing quark masses.

The magnetic moment values in Table~\ref{tab:1} are of similar 
quality as those that are obtained with the static quark model 
in all forms of kinematics.
The results also reveal that the calculated values are notably
closer to the empirical values with finite values for the
constituent masses.
If however, the Dirac magnetic moment operator is employed 
in the naive quark model, the corresponding magnetic moments 
would only be about half as large as the empirical values~\cite{Dannbom}.
The relativistic quark model calculation shows that it
is possible to retain the qualitatively satisfactory
quark model phenomenology of the baryon magnetic moments
even when Dirac currents for the constituent quarks are employed.

In the case of instant form the calculated magnetic moment values 
are almost insensitive to the value of the constituent mass
of the light flavor quark. They are, however, in notably better
agreement with the empirical values when the strange
quark mass is finite and $\sim$ 470 MeV, as shown in
Table \ref{tab:2}. 

The point form values for the hyperon magnetic moments 
given in Tables \ref{tab:1} and \ref{tab:2} are
more sensitive to the value of the constituent mass of the
light flavor quark, but clearly inferior to the instant
form values for all combinations of the quark masses. 

The magnetic moment values that are calculated in front
form are quantitatively fairly similar to those that
are calculated in instant form as shown in Tables \ref{tab:1} and \ref{tab:2}.
With finite quark masses
they fall within 10 \% of the empirical values.   

\begin{table}[t]
\caption{Magnetic moments, in nuclear magnetons, calculated with flavor
independent constituent masses.
The constituent mass used for the $\Omega^-$ is 470 MeV.
The column NR contains the static quark model values
\cite{Dannbom}.\label{tab:1}}
\bce
\begin{tabular}{c|c|c|c|c|c|c|c|c}
Baryon          &  NR & \multicolumn{2}{c|}{Instant}
& \multicolumn{2}{c|}{Point}&  \multicolumn{2}{c}{Front}& EXP \\
\hline
\hline
 ($m_u,m_s$)  & (360,470)  & (0,0)  &  (140,140)     & (0,0) &(350,350)       
   &  (0,0) &(250,250)    &     \\
\hline
\multicolumn{9}{c}{Octet Baryons}\\
\hline
\hline
$p_{(uud)}$         &  2.76    & 2.57     &2.71  & 1.95   & 2.45  & 3.12
  &2.78  & 2.79 \\
$n_{(udd)}$         & $-$1.84    & $-$1.79    &$-$1.84 & $-$1.32  & $-$1.63
 &$-$1.99  &$-$1.67 & $-$1.91 \\
$\Sigma^+_{ (uus)}$ &  2.68    & 2.56     &2.71  & 1.53   & 1.94  & 2.91 
 &2.56  &   2.46 \\
$\Sigma^0_{(uds)}$  &  0.84    & 0.90     &0.92  & 0.52   & 0.65  & 1.00 
 &0.84  &  ?   \\
$\Sigma^-_{(dds)}$  & $-$1.00    & $-$0.77    &$-$0.87 & $-$0.50  &$-$0.64 
 & $-$0.91 &$-$0.89 & $-$1.16\\ 
$\Lambda^0_{(uds)}$ & $-$0.67    & $-$0.90    &$-$0.92 & $-$0.55  &$-$0.69 
 & $-$0.99 &$-$0.84 & $-$0.61 \\
$\Xi^0_{(uss)}$     & $-$1.51    & $-$1.79    &$-$1.84 &  $-$0.94 &$-$1.17 
 & $-$2.00 &$-$1.68 &$-$1.25\\
$\Xi^-_{(dss)}$     & $-$0.59    & $-$0.77    &$-$0.87 & $-$0.45  &$-$0.57 
 &$-$0.83  &$-$0.82 & $-$0.65\\
\hline
\multicolumn{9}{c}{Decuplet Baryons}\\
\hline
\hline
$\Omega^-_{(sss)}$   & $-$2.01    &  $-$2.18  & $-$1.59 &$-$0.93    &$-$1.34 
 & $-$2.09  & $-$1.79 &$-$2.019 \\
$\Delta^{++}_{(uuu)}$&  5.52      &   4.37    &  5.1    & 2.53      & 3.47   
 & 5.38     & 5.21    &4.52\\
\hline
\end{tabular}
\ece
\end{table}

The results that are obtained
with different masses for the $u,d$ and the $s$ quark are shown
in Table~\ref{tab:2}. It is obvious from these results
that the calculated magnetic moment values are overall
much closer to the empirical values, when the constituent
mass of the strange quark is given a value that is
$\sim$ 100 - 200 MeV larger than the constituent mass of the
light flavor quarks. It also emerges that the magnetic
moment values that are obtained with instant and front
form kinematics are closer to the empirical values than
those obtained in point form, when the spatial wave function
of the hyperons is taken to be the same as that for the
nucleon.

\begin{figure}[t]
\vspace{25pt}
\begin{center}
\mbox{\epsfig{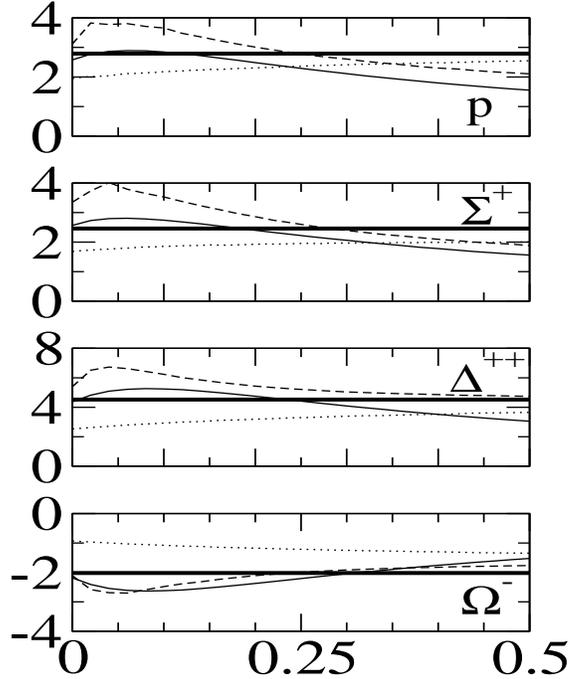}}
\end{center}
\caption{Magnetic moments, in nuclear magnetons, of the 
$p$, $\Sigma^+$ and $\Delta^{++}$ as functions of the 
light constituent quark mass. The mass of the strange 
quark is fixed at 470 MeV for the $\Sigma^+$. 
Magnetic moment, in nuclear magnetons, of the 
$\Omega^-$ as a function of the mass of the strange
quark. Solid, point and dashed lines correspond to instant, 
point and front forms of relativistic kinematics. The 
thick line corresponds
to the experimental value.\label{mmqmass}}
\end{figure}

\begin{figure}[h]
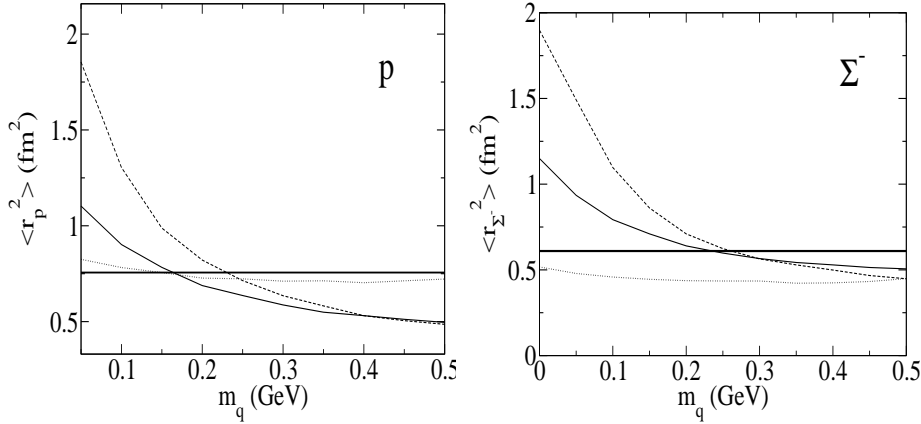

\vspace{25pt}
\begin{center}
\mbox{\epsfig{file=fig2, width=60mm, height=56mm}}
\mbox{\epsfig{file=fig3, width=60mm, height=56mm}}
\end{center}
\caption{Mean square charge radius of the proton and the 
$\Sigma^-$ as a function of the light quark mass, $m_{u,d}$. 
Solid, dotted and dashed lines correspond to instant, point and 
front forms of kinematics. The thick line corresponds
to the experimental value.\label{rpdmq}}
\end{figure}

The dependence of the calculated
magnetic moment of the $p$, $\Sigma^+$, $\Delta^{++}$ 
and the $\Omega^-$ on the value of the constituent quark 
mass is shown in Fig.~\ref{mmqmass}. The figure 
reveals that in point form kinematics there is no value for the
light flavor constituent mass where the calculated
value would agree with the empirical value, whereas
in both instant and in front form kinematics
the curves cross the horizontal line, which
represents the empirical value. Similar but 
clearer features can be seen in Fig.~\ref{rpdmq} where 
the mean square charge radii of the proton and
the $\Sigma^-$ are depicted as  
functions of the light quark mass. In instant and front 
form kinematics there is again a much stronger
dependence on the mass 
of the quark than in point form. 

\begin{table}[t]
\caption{Magnetic moments, in nuclear magnetons, calculated with different masses
for $u,d$ and $s$ quarks.\label{tab:2}}
\bce
\begin{tabular}{c|c|c|c|c|c|c|c}
Baryon          &        
\multicolumn{2}{c|}{Instant} &
\multicolumn{2}{|c|}{Point} &  
 \multicolumn{2}{c|}{Front} &  EXP \\
\hline
\hline
($m_u,m_s$)            & (140,470) & (0,470) & (350,470)&  (0,470
) & (250,470) & (0,470)& \\
\hline
\multicolumn{8}{c}{Octet Baryons}\\
\hline
\hline
$p_{(uud)}$              & 2.71      & 2.57     & 2.45        &1.95       
  & 2.78      &3.12   &  2.79 \\
$n_{(udd)}$              & $-$1.84   &$-$1.80   &  $-$1.63    &$-$1.32   
   & $-$1.67     &$-$1.99  &     $-$1.91 \\
$\Sigma^+_{ (uus)}$      &  2.61     &2.56      &  1.97       &1.69      
   & 2.59      &3.47   &  2.46 \\
$\Sigma^0_{(uds)}$       &  0.79     &0.78      &  0.65       &0.56       
  & 0.75      &0.94   &  ?   \\
$\Sigma^-_{(dds)}$       & $-$1.03   &$-$1.00   &$-$0.67      &$-$0.55    
  &  $-$1.08    &$-$1.60  & $-$1.16\\
$\Lambda^0_{(uds)}$      &  $-$0.55  &$-$0.56   &  $-$0.68    &$-$0.63    
  & $-$0.55     &$-$0.50  & $-$0.61 \\
$\Xi^0_{(uss)}$          &  $-$1.35  &$-$1.37   &$-$1.21      &$-$1.15    
  & $-$1.31     & $-$1.52 &$-$1.25 \\
$\Xi^-_{(dss)}$          &   $-$0.40 &$-$0.38   &$-$0.57      &$-$0.55 
   &  $-$0.59    &$-$0.74  &$-$0.65\\
\hline
\end{tabular}
\ece
\end{table}

Finally Table~\ref{tab:3} contains the calculated values of the 
charge radii of the baryons with different $u,d$ and $s$ 
quark masses. The calculated charge radii for the hyperons
are very similar in instant and front form kinematics.
For the proton, neutron and the $\Sigma^-$ hyperon, for which empirical
values are known, the calculated values are also close to the empirical
values. In the case of the neutron, agreement with the empirical
value does however require the introduction of a mixed symmetry $S-$state
component into the ground state wave function, with a norm of 0.01 - 0.02.

The point form value for the mean square charge radius of the
$\Sigma^-$ is considerably smaller than the empirical
value. This is clearly related to the fact that in
point form kinematics the magnetic moment of the
$\Sigma^-$ also is only about half of the experimental
value. The same conclusion applies to the case
of the calculated charge radius and magnetic
moment of the $\Omega^-$ hyperon. This implies that
in point form kinematics, the spatial wave function
of the hyperons has to differ significantly from that
of the nucleons. In order to get the empirical value for the
mean square charge radius of the $\Sigma^-$ hyperon in point
form kinematics the parameter $b$ in the spatial wave function 
(\ref{ground}) has to be decreased from the value 640 MeV
in Table \ref{tab:1} to 300 MeV, which corresponds to more
than doubling the matter radius. This increases both the
calculated mean square radii and the magnetic moments of most
of the strange hyperons to agree better with both the
empirical results and the results obtained with the two other
forms of kinematics. This is shown in Table~\ref{tab:new}.

\begin{table}[t]
\caption{Mean square charge radii in fm$^2$ calculated 
with different $u,d$ and $s$ quark masses. The constituent 
mass used for the $\Omega^-$ is 470 MeV.
For the neutron the values obtained with a small mixed 
symmetry$-S$ state component in the
wave function as in Ref.~\cite{Bruno} are shown in brackets.\label{tab:3}}
\bce
\begin{tabular}{c|c|c|c|c}
Baryon          &  \multicolumn{1}{c|}{Instant}
& \multicolumn{1}{c|}{Point}&  \multicolumn{1}{c|}{Front}& EXP \\
\hline
\hline
 ($m_u,m_s$)         &    (140,470)     &(350,470)      &(250,470) 
   &     \\
\hline
\multicolumn{5}{c}{Octet Baryons}\\
\hline
\hline
$p_{(uud)}$         &0.79     & 0.71    & 0.71    & 0.756~\protect\cite{PDG}  \\
$n_{(udd)}$         &0.00($-$0.14)& $-$0.01 ($-$0.13) & $-$0.02 ($-$0.09)
 & $-$0.116~\protect\cite{PDG}      \\
$\Sigma^+_{ (uus)}$ &1.10     & 0.47     & 0.94     &       \\  
$\Sigma^0_{(uds)}$  &0.19     & 0.02     & 0.13     &       \\ 
$\Sigma^-_{(dds)}$  &0.72  & 0.41  & 0.61  & 0.61
~\protect\cite{sigman}     \\
$\Lambda^0_{(uds)}$ &0.19     & $-$0.01  & 0.12     &       \\
$\Xi^0_{(uss)}$     &0.34     & 0.01     & 0.22     &       \\
$\Xi^-_{(dss)}$     &0.61  & 0.35  & 0.57  &       \\
\hline
\multicolumn{5}{c}{Decuplet Baryons}\\
\hline
\hline
$\Omega^-_{(sss)}$  &   0.54      &   0.25       & 0.51  &  \\
$\Delta^{++}_{(uuu)}$&  1.64       &    0.85        &  1.47    & \\
\hline
\end{tabular}
\ece
\end{table}

\begin{table}[t]
\caption{Magnetic moments, in nuclear magnetons, and mean 
square charge radii, in fm$^2$, calculated with 
$m_{u,d}=350$ MeV and $m_s=470$ MeV using point form of 
relativistic kinematics. The results are obtained making use 
of two different values for the oscillator parameter 
$b$.\label{tab:new} }
\bce
\begin{tabular}{c|c|c|c}
Baryon          &        
\multicolumn{1}{c|}{($b$=640)} &
\multicolumn{1}{|c|}{($b$=300)} &  EXP \\
\hline
\multicolumn{4}{c}{Magnetic moments}\\
\hline
\hline
$\Sigma^+_{ (uus)}$      &  1.97       & 2.20     &  2.46 \\
$\Sigma^0_{(uds)}$       &  0.65       & 0.71     &  ?   \\
$\Sigma^-_{(dds)}$       &$-$0.67      &  $-$0.77 & $-$1.16\\
$\Lambda^0_{(uds)}$      &  $-$0.68    & $-$0.69  & $-$0.61 \\
$\Xi^0_{(uss)}$          &$-$1.21      & $-$1.31  &$-$1.25 \\
$\Xi^-_{(dss)}$          &$-$0.57      &  $-$0.58 &$-$0.65\\
\hline
\multicolumn{4}{c}{Mean square charge radii} \\
\hline
$\Sigma^+_{ (uus)}$ & 0.47     & 0.66     &       \\  
$\Sigma^0_{(uds)}$  & 0.02     & 0.02     &       \\ 
$\Sigma^-_{(dds)}$  & 0.41  & 0.64  & 0.61
~\protect\cite{sigman}     \\
$\Lambda^0_{(uds)}$ & $-$0.01  & 0.02     &       \\
$\Xi^0_{(uss)}$     & 0.01     & 0.03     &       \\
$\Xi^-_{(dss)}$     & 0.35     & 0.52  &       \\
\hline
\end{tabular}
\ece
\end{table}

\section{Magnetic Moments and Charge Radii of the Charm Hyperons}

\subsection{Charm 1 Hyperons}

The $SU(3)_F$ quark model may be directly extended to 
$SU(4)_F$ to encompass the charm baryons. The flavor
symmetry breaking is in this case large because of the 
large difference between the charm and light flavor
constituent quark masses. The main flavor symmetry breaking
appears through the quark currents and
the boosts and spin rotations, which depend explicitly on the
constituent mass.
The flavor symmetry breaking in the mass operator may either
be treated perturbatively or
in the form of a flavor and spin dependent 
hyperfine term, which is independent of the spatial
coordinates~\cite{codann}. This makes it possible to
employ a flavor independent spatial wave function and
to use the same spatial wave function for the charm
and strange hyperons.

For $SU(4)$ symmetry there are two 20plets, one of which
corresponds to mixed symmetry states and the other 
one to the completely symmetric flavor wave states. The 
completely symmetric flavor states represent spin 3/2 states, 
while the mixed symmetry ones represent spin 1/2 states.

The wave functions for the mixed symmetry 20plet are 
listed in Table~\ref{bwf4}. The corresponding wave 
functions for the symmetric 20plet are readily constructed 
by symmetrization of the quark content of the baryon.

The spin-flavor matrix elements that enter in the calculation of 
the magnetic moments and charge radii of the charm
hyperons may, in analogy with the case of the strange
hyperons above, be written as
$\bra{B}\,  {\mathcal S \mathcal F}_c \,  \ket{B'}$.
Here the flavor operator ${\mathcal F}_c$ is the extension to charm
of the charge operator (\ref{charge}):
\beq
 {\mathcal F}_c = \bigg\{ {\lambda_3 \over 2} + 
{\lambda_8\over 2\sqrt{3}} + {1\over 6}(1-\sqrt{6}\lambda_{15}) \bigg\} \, .
\label{chargec}
\eeq
The explicit expressions for the matrix elements of the
flavor operator as obtained with the wave functions in Table~\ref{bwf4}
are given in Appendix~\ref{ap:me}.
These expressions reproduce the static results~\cite{lichten} 
when the spin operator is approximated
by the static operator of Eq.~(\ref{nonrel}). 
In the relativistic case the matrix elements of the quark 
operator depend on the form of kinematics under consideration.

Tables~\ref{tab:6} and~\ref{tab:7} contain the calculated
values of the magnetic 
moments of charm baryons with mixed symmetry for the three
different forms of kinematics. 
The results in Table~\ref{tab:6} are given for both zero and
finite values for the mass of the light quarks, while the
strange and charm quark are given the constituent masses
470 MeV and 1500 MeV respectively. In instant and point form
kinematics the results are insensitive to the value
of the constituent mass of the light flavor quark, whereas
in front form the calculated values are quite sensitive to
the light quark mass. When the light flavor quarks have
finite constituent masses the results are very similar for
all three forms of kinematics. 

To show the sensitivity of the
calculated magnetic moments of the charm hyperons to the
value of the constituent mass of the strange quark, the
results that are obtained by setting the mass of the strange
quark, along with those of the light flavor quarks, to zero are 
shown in Table~\ref{tab:7}. We therefore employ a finite
value for the mass of the charm and zero mass for the
$u,d$ and $s$ quarks. Both instant and front form values
reveal notable sensitivity
to the value of the strange quark mass, whereas the point
form values are again very insensitive to the strange
quark mass.

\begin{table}[t]
\caption{Magnetic moments, in nuclear magnetons, of the charm hyperons
calculated with different masses for $u,d$,
$s$ and $c$ quarks for the three forms of kinematics. 
The strange mass used is 470 MeV in the three cases. The charm quark 
mass is taken to be 1500 MeV. Here $M_{\Xi_{cc}^{++}} = M_{\Xi_{cc}^{+}}$=3.5 GeV
\cite{selex1} and $M_{\Omega_{cc}^{+}}$= 3.7 GeV.
The column NR shows the corresponding
values in the static quark model with these mass values
and $m_u=m_d=$350 MeV. \label{tab:6}}
\bce
\begin{tabular}{c|c|c|c|c|c|c|c}
Baryon          &        
\multicolumn{2}{c|}{Instant} &
\multicolumn{2}{|c|}{Point} &  
 \multicolumn{2}{c|}{Front} & NR \\
\hline
\hline
($m_u$)         & 140   & 0   & 350  &  0  & 250  & 0    & \\
\hline
\multicolumn{8}{c}{Mixed symmetry 20plet: (6,1)}\\
\hline
\hline
$\Sigma_c^{++}(uuc)$ & 2.36   &   2.39  &  1.00  &  0.90   &2.18 &3.07 & 2.24   \\
$\Sigma_c^+(udc)$  & 0.49   &   0.49  &  0.13  &  0.11   &0.44 & 0.65& 0.46   \\ 
$\Sigma_c^0(ddc)$  & $-$1.38  &  $-$1.40 & $-$0.74 &  $-$0.67 &$-$1.31 & $-$1.78& $-$1.33  \\
$\Xi_c^{s+}(usc)$  &  0.75  &  0.78    &   0.15  &  0.13    &0.67 & 1.13& 0.61   \\
$\Xi_c^{s0}(dsc)$  &$-$1.12 & $-$1.14  &$-$0.73  & $-$0.70  &$-$1.24 & $-$1.51& $-$1.18 \\
$\Omega^{0}_c(ssc)$&$-$0.86 &  $-$0.86 &$-$0.67  &  $-$0.67 &$-$0.90 & $-$0.90& $-$1.03 \\
\hline
\multicolumn{8}{c}{Mixed symmetry 20plet: ($\bar{3}$,1)} \\
\hline
$\Xi_c^{a+}(usc)$  & 0.40      &  0.39    &   0.47  &   0.45   &0.40 &0.39 &  0.42  \\
$\Xi_c^{a0}(dsc)$  & 0.41      &  0.42    &   0.47  &   0.46   &0.40 &0.39 & 0.42   \\ 
$\Lambda_c^+(udc)$ &0.40       & 0.39      & 0.52   &   0.49   &0.43 &  0.41 & 0.42   \\
\hline
\multicolumn{8}{c}{Mixed symmetry 20plet: (3,2)} \\
\hline
$\Xi_{cc}^{++}(ucc)$& $-$0.10  & $-$0.13 & 0.30  & 0.29 & $-$0.04 & $-$0.32 & $-$0.04 \\
$\Xi_{cc}^{+}(dcc)$ &    0.86  & 0.87    & 0.69  & 0.68 & 0.83    & 0.94  & 0.85\\
$\Omega_{cc}^{+}(scc)$ & 0.72  & 0.72    & 0.66  & 0.66 & 0.74    & 0.74   & 0.78\\  
\hline
\end{tabular}
\ece
\end{table}

\begin{table}[t]
\caption{Magnetic moments, in nuclear magnetons, calculated with zero mass 
for the light quarks, $u,d$ and $s$ and with the charm quark mass
as 1500 MeV. \label{tab:7}}
\bce
\begin{tabular}{c|c|c|c}
Baryon          &        
\multicolumn{1}{c|}{Instant} &
\multicolumn{1}{|c|}{Point} &  
 \multicolumn{1}{c}{Front}  \\
\hline
\hline
\multicolumn{4}{c}{Mixed symmetry 20plet: (6,1)}\\
\hline
\hline
$\Xi_c^{s+}(usc)$    &0.49    &  0.11  & 0.64\\
$\Xi_c^{s0}(dsc)$    &$-$1.40 & $-$0.67&$-$ 1.96\\
$\Omega^{0}_c(ssc)$  &$-$1.40 & $-$0.61&$-$1.77 \\
\hline
\multicolumn{4}{c}{Mixed symmetry 20plet: ($\bar{3}$,1)} \\
\hline
$\Xi_c^{a+}(usc)$    &0.39    &  0.43  & 0.36\\
$\Xi_c^{a0}(dsc)$    &0.43    &  0.45  & 0.35\\
\hline
\multicolumn{4}{c}{Mixed symmetry 20plet: (3,2)} \\
\hline
$\Omega_{cc}^{+}(scc)$ & 0.87    &  0.65  & 0.94 \\
\hline
\end{tabular}
\ece
\end{table}

In the absence of empirical data on the magnetic moments
of the charm hyperons, the values obtained by the static
quark model with the corresponding values are given
as a reference. The calculated values in instant form
kinematics are very similar to the static quark model
values, independently of the value of the constituent
mass of the light flavor quarks. This similarity also
holds in the case of front form kinematics, although
only if the constituent mass of the light flavor
quarks is finite.

As in the case of the strange hyperons,
point form kinematics does, however, give much smaller 
values for the magnetic moments of the charm hyperons than
both the instant and front form kinematics.
The reason for this is that the large hyperon mass
appears in the boost velocity in the case of point
form kinematics. The larger
values are expected to be the more realistic ones, however,
especially in view of the fact that those values are also similar
to the values given by the extension to charm of the 
Skyrme model \cite{scoccola}. 

Table~\ref{tab:8} contains the calculated values for the 
mean square charge radii that are obtained with different masses
for the light, strange and charmed quarks. The 
point form results again deviate considerably from the
instant and front form results, which are similar to one another. 
In particular the charge radii 
obtained for the neutral hyperons in point form 
are usually much smaller than the those obtained with instant 
and front forms of kinematics. This feature is already 
noticeable in the results of Table~\ref{tab:3} for the  
strange baryons.

\begin{table}[t]
\caption{Mean square charge radii (in fm$^2$) of the charm
hyperons calculated with 
different masses for quarks $u, d$, $s$ and $c$. The masses of 
the strange and charm quarks, $m_s$ and $m_c$, are taken to be 470 MeV
and 1.5 GeV respectively.
\label{tab:8}}
\bce
\begin{tabular}{c|c|c|c}
Baryon          &        
\multicolumn{1}{c|}{Instant} &
\multicolumn{1}{|c|}{Point} &  
 \multicolumn{1}{c}{Front}  \\
\hline
\hline
\multicolumn{4}{c}{Mixed symmetry 20plet: (6,1)}\\
\hline
\hline
$\Sigma^{++}(uuc)$     & 1.7    & 0.4    &1.4     \\
$\Sigma_c^+(udc)$      & 0.5    & 0.2    & 0.4    \\
$\Sigma_c^0(ddc)$      &$-$0.7  &$-$0.0  & $-$0.6 \\
$\Xi_c^{s+}(usc)$      &  0.6   & 0.2    & 0.5    \\
$\Xi_c^{s0}(dsc)$      &$-$0.6  & $-$0.0  &$-$0.5  \\
$\Omega^{0}_c(ssc)$    &$-$0.4  & $-$0.0  &$-$0.4  \\
\hline
\multicolumn{4}{c}{Mixed symmetry 20plet: ($\bar{3}$,1)} \\
\hline
$\Xi_c^{a+}(usc)$      & 0.6    & 0.2    &0.5     \\
$\Xi_c^{a0}(dsc)$      & $-$0.6 & $-$0.0 &$-$0.5  \\
$\Lambda_c^+(udc)$     & 0.5    & 0.2    & 0.4    \\
\hline
\multicolumn{4}{c}{Mixed symmetry 20plet: (3,2)} \\
\hline
$\Xi_{cc}^{++}(ucc)$   & 1.3     & 0.2    & 1.0      \\
$\Xi_{cc}^{+}(dcc)$    & 0.1     & 0.1    & 0.0  \\
$\Omega_{cc}^{+}(scc)$ & 0.2     & 0.1    & 0.1  \\
\hline
\end{tabular}
\ece
\end{table}

In the case of the strange hyperons it was found that
acceptable values of the calculated magnetic moments
and mean square charge radii are obtained in point
form kinematics only if the parameter $b$ in the
spatial wave function model (\ref{ground}) is reduced
from 640 MeV to 300 MeV. This also applies to the
case of the charm hyperons as shown in  
Tables~\ref{tab:new2} and~\ref{tab:new3}, where the calculated magnetic
moments and mean square charge radii are shown for
both $b=$640 MeV and $b=300$ MeV with point form
kinematics. The substantial
reduction of the parameter $b$ again leads 
values for both observables, which are much closer to
the values that are obtained with instant and front
form kinematics.

\begin{table}[h]
\caption{Magnetic moments, in nuclear magnetons, of the charm hyperons
calculated in point form with $m_{u,d}=350$ MeV, $m_s=470$ MeV and $m_c=1.5$ GeV.
Here $M_{\Xi_{cc}^{++}} = M_{\Xi_{cc}^{+}}$=3.5 GeV
\cite{selex1} and $M_{\Omega_{cc}^{+}}$= 3.7 GeV.
The column NR shows the corresponding
values in the static quark model with these mass values.~\label{tab:new2}}
\bce
\begin{tabular}{c|c|c|c}
Baryon          &        
\multicolumn{1}{c|}{ ($b$=640)} &
\multicolumn{1}{c|}{($b$=300)} &   NR \\
\hline
\hline
\multicolumn{4}{c}{Mixed symmetry 20plet: (6,1)}\\
\hline
\hline
$\Sigma_c^{++}(uuc)$ &  1.00  &  1.39   & 2.24   \\
$\Sigma_c^+(udc)$    &  0.13  &   0.24  & 0.46   \\ 
$\Sigma_c^0(ddc)$  & $-$0.74  &$-$0.91  & $-$1.33  \\
$\Xi_c^{s+}(usc)$  &   0.15   &0.30     & 0.61   \\
$\Xi_c^{s0}(dsc)$  &$-$0.73   &$-$0.88  & $-$1.18 \\
$\Omega^{0}_c(ssc)$&$-$0.67   &$-$0.78  & $-$1.03 \\
\hline
\multicolumn{4}{c}{Mixed symmetry 20plet: ($\bar{3}$,1)} \\
\hline
$\Xi_c^{a+}(usc)$  &   0.47  &0.42 &  0.42  \\
$\Xi_c^{a0}(dsc)$  &   0.47  &0.42 & 0.42   \\ 
$\Lambda_c^+(udc)$ &   0.52  &0.46 & 0.42   \\
\hline
\multicolumn{4}{c}{Mixed symmetry 20plet: (3,2)} \\
\hline
$\Xi_{cc}^{++}(ucc)$& 0.30  & 0.17     & $-$0.04 \\
$\Xi_{cc}^{+}(dcc)$ & 0.69  & 0.74       & 0.85\\
$\Omega_{cc}^{+}(scc)$ & 0.66  & 0.69    & 0.78\\  
\hline
\end{tabular}
\ece
\end{table}

\begin{table}[h]
\caption{Mean square charge radii (in fm$^2$) of the charm
hyperons calculated in point form with $m_{u,d}=350$ MeV, 
$m_s=470$ MeV and $m_c=1500$ MeV.\label{tab:new3}}
\bce
\begin{tabular}{c|c|c}
Baryon          &        
\multicolumn{1}{c|}{ ($b$=640)} &
\multicolumn{1}{|c}{($b$=300) } \\
\hline
\multicolumn{3}{c}{Mixed symmetry 20plet: (6,1)}\\
\hline
\hline
$\Sigma^{++}(uuc)$     & 0.4    &1.1     \\
$\Sigma_c^+(udc)$      & 0.2    & 0.5    \\
$\Sigma_c^0(ddc)$      &$-$0.0  & $-$0.1 \\
$\Xi_c^{s+}(usc)$      & 0.2    & 0.5    \\
$\Xi_c^{s0}(dsc)$      & $-$0.0  &$-$0.1  \\
$\Omega^{0}_c(ssc)$    & $-$0.0  &$-$0.0  \\
\hline
\multicolumn{3}{c}{Mixed symmetry 20plet: ($\bar{3}$,1)} \\
\hline
$\Xi_c^{a+}(usc)$      & 0.2    &0.5     \\
$\Xi_c^{a0}(dsc)$      & $-$0.0 &$-$0.0  \\
$\Lambda_c^+(udc)$     & 0.2    & 0.6    \\
\hline
\multicolumn{3}{c}{Mixed symmetry 20plet: (3,2)} \\
\hline
$\Xi_{cc}^{++}(ucc)$   & 0.2    & 0.5      \\
$\Xi_{cc}^{+}(dcc)$    & 0.1    & 0.2  \\
$\Omega_{cc}^{+}(scc)$ & 0.1    & 0.2  \\
\hline
\end{tabular}
\ece
\end{table}

\subsection{Doubly Charmed Hyperons}

The recent discovery of doubly charmed hyperons, with rest energies
at 3460, 3520 and 3780 MeV has revived the interest in the structure
of those baryons \cite{selex1,selex2}. It is natural to identify the
state at 3460 MeV as the $\Xi_{cc}^+$, pending verification that its spin
and parity assignment be $1/2^+$. The state at 3520 MeV may most
naturally be identified as the $\Xi_{cc}^{*++}$ in view of the
prediction that the splitting between these states should be
expected to be $\sim 60 - 90$ MeV 
\cite{codann,richard,rho,itoh,woloshyn}.
The state at 3780 MeV is likely to be the lowest negative parity
state, in view of the prediction that this should fall around
300 MeV above the ground state \cite{richard}.

The magnetic moments of the ground states of the doubly charmed hyperons 
may be calculated by the same methods as used above for the strange
and singly charmed hyperons. In Table \ref{tab:6} the calculated
magnetic moments of the ground state $ccu$ ($\Xi_{cc}^{++}$),
$ccd$ ($\Xi_{cc}^{+}$) and $ccs$ ($\Omega_{cc}^{+}$) are given
as obtained in the three different forms of kinematics. Here
the mass of the $\Xi_{cc}^{++}$ has been taken as 3.5 GeV
and that of the $\Omega_{cc}^+$ as 3.7 MeV. The latter value
is suggested by the lattice calculation in Ref.~\cite{woloshyn}.
For reference we note that the static quark model expressions for the
magnetic moments (in units of nuclear magnetons) of these hyperons are:
\beqa
 \mu(\Xi_{cc}^{++}) &=& {8\over 9}{m_p\over m_c} - {2\over 9} {m_p\over m_u}\, ,
\nonumber \\ 
 \mu(\Xi_{cc}^{+})  &=& {8\over 9}{m_p\over m_c} + {1\over 9} {m_p\over m_u}\, ,
\nonumber \\ 
 \mu(\Omega_{cc}^{+}) &=& {8\over 9}{m_p\over m_c} + {1\over 9} {m_p\over m_s}\, . 
\label{dblchmag}
\eeqa
Here $m_u$, $m_s$ and $m_c$ are the constituent masses of the light flavor,
the strange and the charm quarks respectively.

The calculated magnetic moments of the doubly charmed hyperons are very
similar in instant and in front form kinematics. 
They are also close to the static quark model values.
In contrast the
point form values differ notably from these, in particular
in the case of
the $\Xi_{cc}^{+}$, for which the point 
form result is positive while both instant 
and front give a negative result for the magnetic moment.

In Table \ref{tab:7} the calculated result for the magnetic moment of the 
$\Omega_{cc}^+$ is also shown for the case of zero constituent
mass for the strange quark. Only the front form result shows 
significant dependence on the quark mass, in analogy with the
case of the strange hyperons. 

In Table \ref{tab:8} the calculated mean square charge radii of the
ground state doubly charmed hyperons are listed. As in the case of
the magnetic moments, the instant and front form values are
similar, while the point form values are much smaller. 

\section{Discussion}

The main finding above is the demonstration that the phenomenological
description of baryon magnetic moments may be carried over into the
Poincar\'e
covariant quark model, and with the additional gain of a concomitant
description of the charge radii, and, for the nucleons and the
$\Delta(1232)$ resonance, of the electromagnetic form factors as well.
In comparison with the static quark model the gain is the dynamical
consistency of the calculation. Formally the calculation of the current 
matrix elements is considerably more cumbersome in the relativistic
quark model, the results of which coincide with those of the static
quark model only when $m_q/m_n \gg 1$. 

Another notable finding is that if the same spatial wave function
is employed the strange hyperon magnetic moments and mean square
charge radii calculated
in point 
form kinematics are considerably farther from the empirical magnetic
moments than the
values calculated in instant and front form kinematics. Both instant
and front
form kinematics lead to very similar values for the hyperon magnetic
moments,
and moreover to values, which are close to the empirical values. 
Only if the spatial extent of the wave function of the strange
hyperons is more than double that of the spatial extent of the
nucleon wave function does the calculation in point form kinematics
lead to values that are close to the corresponding empirical ones.
A similar conclusion applies to the case of the charm
hyperons, for which point form kinematics in most cases yields
much smaller 
magnetic moment values than instant and front form unless
the spatial extent of the wave function is more than
double that of the nucleons. That point form kinematics 
differs qualitatively from the other two forms has
been noted in another context in Ref.~\cite{despl}.

Finally the present results suggest that the apparent success of the 
static 
quark model with constituent masses in the range 300 $-$ 400 MeV in
describing the
empirical baryon magnetic moments with $\sim$ 10 $-$ 15\% accuracy is
largely an accidental consequence of the correct description of the
flavor structure of current operators. 

The present phenomenlogical approach to the baryon magnetic
moments has no direct link to QCD. There are several QCD 
based approaches to the observables of the baryons in the
low-energy range. These are based on effective
field theory \cite{meissner,durand} or on the large color 
limit \cite{jenkins1,jenkins} and in both cases lead to a
description in terms of effective operators. The issue
of Poincar\'e covariance may, to a good approximation,
be avoided in leading order in the effective operator expansion.

\section*{Acknowledgments}

We are grateful for patient instruction by F. Coester.
D. O. R. thanks R. D. McKeown for hospitality at the
W. K. Kellogg Radiation Laboratory at the California
Institute of Technology during the completion of this
work.
Research supported in part by the Academy of Finland through
grant 54038 and the European Euridice network
HPRN-CT-2002-00311.


\appendix

\section{Flavor matrix elements}
\label{ap:me}

Here the explicit expressions for the spin-flavor
matrix elements that enter in the evaluation of 
both the magnetic moments and the charge radii are given.
These matrix elements arise upon calculation
of the matrix element of the flavor operator with the
wave functions of Table \ref{bwf4}.

\subsection{Non-strange non-charm sector}

The expressions for the octet baryons are,
\beqa
\bra{p} \; {\mathcal S F} \; \ket{p}  &=& {1\over 3} \bra{S_{MA}} \, {\mathcal S}\,
 \ket{S_{MA}}_u  \label{SI} \\
\bra{n} \; {\mathcal S F} \; \ket{n}  &=& {1\over 6}\bra{S_{MS}} \, {\mathcal S}\,
 \ket{S_{MS}}_u -
                                      {1\over 6} \bra{S_{MA}} \, {\mathcal S}\,
 \ket{S_{MA}}_u \nonumber \\
\bra{\Sigma^+} \; {\mathcal S F} \; \ket{\Sigma^+}&=& 
  {1\over3} \bra{S_{MA}} \, {\mathcal S}\, \ket{S_{MA}}_u 
+ {1\over9} \bra{S_{MS}} \, {\mathcal S}\, \ket{S_{MS}}_u \nonumber \\
&&- {1\over 9}\bra{S_{MS}} \, {\mathcal S}\, \ket{S_{MS}}_s
\nonumber \\
\bra{\Sigma^0} \; {\mathcal S F} \; \ket{\Sigma^0}&=&    
  {1\over 12} \bra{S_{MA}} \, {\mathcal S}\, \ket{S_{MA}}_u 
+ {1\over 36} \bra{S_{MS}} \, {\mathcal S}\, \ket{S_{MS}}_u \nonumber \\ 
&&- {1\over 9}  \bra{S_{MS}} \, {\mathcal S}\, \ket{S_{MS}}_s
\nonumber \\
\bra{\Sigma^-} \; {\mathcal S F} \; \ket{\Sigma^-}&=&    
-{1\over6}   \bra{S_{MA}} \, {\mathcal S}\, \ket{S_{MA}}_u 
-{1\over 18} \bra{S_{MS}} \, {\mathcal S}\, \ket{S_{MS}}_u \nonumber \\
&&-{1\over 9}  \bra{S_{MS}} \, {\mathcal S}\, \ket{S_{MS}}_s
\nonumber \\
\bra{\Lambda^0} \; {\mathcal S F} \; \ket{\Lambda^0}&=&    
{1\over36}   \bra{S_{MA}} \, {\mathcal S}\, \ket{S_{MA}}_u 
+ {1\over12} \bra{S_{MS}} \, {\mathcal S}\, \ket{S_{MS}}_u \nonumber \\
&&- {1\over 9}\bra{S_{MA}} \, {\mathcal S}\, \ket{S_{MA}}_s
\nonumber \\
\bra{\Xi^-} \; {\mathcal S F} \; \ket{\Xi^-}&=&    
-{1\over6}  \bra{S_{MA}} \, {\mathcal S}\, \ket{S_{MA}}_s 
-{1\over18} \bra{S_{MS}} \, {\mathcal S}\, \ket{S_{MS}}_s \nonumber \\
&&-{1\over 9} \bra{S_{MS}} \, {\mathcal S}\, \ket{S_{MS}}_u
\nonumber \\
\bra{\Xi^0} \; {\mathcal S F} \; \ket{\Xi^0}&=&    
-{1\over6}  \bra{S_{MA}} \, {\mathcal S}\, \ket{S_{MA}}_s 
-{1\over18} \bra{S_{MS}} \, {\mathcal S}\, \ket{S_{MS}}_s \nonumber \\
&&+{2\over 9} \bra{S_{MS}} \, {\mathcal S}\, \ket{S_{MS}}_u 
\nonumber 
\eeqa
For the decuplet ones,
\beqa
\bra{\Delta^{++}} \; {\mathcal S F} \; \ket{\Delta^{++}}&=&    
2 \; \bra{S_S} \, {\mathcal S} \, \ket{S_S}_u \nonumber \\
\bra{\Omega^{-}} \; {\mathcal S F} \; \ket{\Omega^{-}}&=&    
(-1) \;  \bra{S_S} \, {\mathcal S} \, \ket{S_S}_s
\eeqa
where for instance, $\bra{S_{MA}} \, {\mathcal S}\, \ket{S_{MA}}_s$, stands
for a matrix element between two mixed antisymmetric states of the 
spin operator ${\mathcal S}$ where the struck quark is a strange quark. 
Depending on the baryon the other two quarks could be $uu$ ($\Sigma$) or
$us$ ($\Xi$). 

\subsection{Strange and charm sector of the mixed symmetry 20plet}

The elastic matrix elements required for the 20plet (6,1) are:
\beqa
\bra{\Sigma_c^{++}} \; {\mathcal S F} \; \ket{\Sigma_c^{++}}&=& 
   {1\over 3} \bra{S_{MA}} \, {\mathcal S}\, \ket{S_{MA}}_u 
 + {1\over 9} \bra{S_{MS}} \, {\mathcal S}\, \ket{S_{MS}}_u \nonumber \\
&& + {2\over 9} \bra{S_{MS}} \, {\mathcal S}\, \ket{S_{MS}}_c
 \nonumber \\ 
\bra{\Sigma_c^{+}} \; {\mathcal S F} \; \ket{\Sigma_c^{+}}&=& 
 {1\over 36}\bra{S_{MS}} \, {\mathcal S}\, \ket{S_{MS}}_u 
+{4\over 18}\bra{S_{MS}} \, {\mathcal S}\, \ket{S_{MS}}_c  \nonumber \\
&&+{1\over12} \bra{S_{MA}} \, {\mathcal S}\, \ket{S_{MA}}_u
 \nonumber \\
\bra{\Sigma_c^{0}} \; {\mathcal S F} \; \ket{\Sigma_c^{0}}&=& 
 -{1\over 18}  \bra{S_{MS}} \, {\mathcal S}\, \ket{S_{MS}}_u 
 +{4\over 18}  \bra{S_{MS}} \, {\mathcal S}\, \ket{S_{MS}}_c  \nonumber \\
&& -{1\over 6}   \bra{S_{MA}} \, {\mathcal S}\, \ket{S_{MA}}_u 
\nonumber \\
\bra{\Xi^+_c} \; {\mathcal S F} \; \ket{\Xi^+_c}&=& 
  {1\over 18}  \bra{S_{MS}} \, {\mathcal S}\, \ket{S_{MS}}_u 
 -{1\over 36}  \bra{S_{MS}} \, {\mathcal S}\, \ket{S_{MS}}_s  \nonumber \\
&& +{4\over 18}  \bra{S_{MS}} \, {\mathcal S}\, \ket{S_{MS}}_c 
 +{1\over 6}  \bra{S_{MA}} \, {\mathcal S}\, \ket{S_{MA}}_u  \nonumber \\
&&-{1\over 12}  \bra{S_{MA}} \, {\mathcal S}\, \ket{S_{MA}}_s 
\nonumber \\
\bra{\Xi^0_c} \; {\mathcal S F} \; \ket{\Xi^0_c}&=& 
 -{1\over 36}  \bra{S_{MS}} \, {\mathcal S}\, \ket{S_{MS}}_u 
 -{1\over 36}  \bra{S_{MS}} \, {\mathcal S}\, \ket{S_{MS}}_s  \nonumber \\
&& +{4\over 18}  \bra{S_{MS}} \, {\mathcal S}\, \ket{S_{MS}}_c 
 -{1\over 12}  \bra{S_{MA}} \, {\mathcal S}\, \ket{S_{MA}}_u  \nonumber \\
&& -{1\over 12}  \bra{S_{MA}} \, {\mathcal S}\, \ket{S_{MA}}_s 
\nonumber \\
\bra{\Omega^0_c} \; {\mathcal S F} \; \ket{\Omega^0_c}&=&
 -{1\over 18}  \bra{S_{MS}} \, {\mathcal S}\, \ket{S_{MS}}_s 
 +{4\over 18}  \bra{S_{MS}} \, {\mathcal S}\, \ket{S_{MS}}_c \nonumber \\
&& -{1\over 6}   \bra{S_{MA}} \, {\mathcal S}\, \ket{S_{MA}}_s
\eeqa
For the 20plet($\bar{3}$,1) we get,
\beqa
\bra{\Xi^{+}_{c\bar{3}}} \; {\mathcal S F} \; \ket{\Xi^{+}_{c\bar{3}}}&=&
  {1\over 6}  \bra{S_{MS}} \, {\mathcal S}\, \ket{S_{MS}}_u 
 -{1\over 12}  \bra{S_{MS}} \, {\mathcal S}\, \ket{S_{MS}}_s \nonumber \\
&&
 +{1\over 18}  \bra{S_{MA}} \, {\mathcal S}\, \ket{S_{MA}}_u 
 -{1\over 36}  \bra{S_{MA}} \, {\mathcal S}\, \ket{S_{MA}}_s  \nonumber \\
&& +{4\over 18}  \bra{S_{MA}} \, {\mathcal S}\, \ket{S_{MA}}_c 
\nonumber \\
\bra{\Xi^{0}_{c\bar{3}}} \; {\mathcal S F} \; \ket{\Xi^{0}_{c\bar{3}}}&=&
 - {1\over 12}  \bra{S_{MS}} \, {\mathcal S}\, \ket{S_{MS}}_u 
 -{1\over 12}  \bra{S_{MS}} \, {\mathcal S}\, \ket{S_{MS}}_s  \nonumber \\
&&
 -{1\over 36}  \bra{S_{MA}} \, {\mathcal S}\, \ket{S_{MA}}_u 
 -{1\over 36}  \bra{S_{MA}} \, {\mathcal S}\, \ket{S_{MA}}_s  \nonumber \\
&&+{4\over 18}  \bra{S_{MA}} \, {\mathcal S}\, \ket{S_{MA}}_c 
\nonumber \\
\bra{\Lambda^{+}_{c\bar{3}} } \; {\mathcal S F} \; \ket{\Lambda^{+}_{c\bar{3}} }&=&
  {1\over 12}  \bra{S_{MS}} \, {\mathcal S}\, \ket{S_{MS}}_u 
 +{1\over 36}  \bra{S_{MA}} \, {\mathcal S}\, \ket{S_{MA}}_u \nonumber \\
&& +{4\over 18}  \bra{S_{MA}} \, {\mathcal S}\, \ket{S_{MA}}_c \, .
\eeqa

For the 20plet(3,2) the matrix elements are:
\beqa
\bra{\Xi^{++}_{cc} } \; {\mathcal S F} \; \ket{\Xi^{++}_{cc}}&=&
  {4\over 18}  \bra{S_{MS}} \, {\mathcal S}\, \ket{S_{MS}}_u 
 +{2\over 18}  \bra{S_{MS}} \, {\mathcal S}\, \ket{S_{MS}}_c  \nonumber \\
&& +{1\over 3}  \bra{S_{MA}} \, {\mathcal S}\, \ket{S_{MA}}_c
\nonumber \\
\bra{\Xi^{+}_{cc} } \; {\mathcal S F} \; \ket{\Xi^{+}_{cc}}&=&
 -{2\over 18}  \bra{S_{MS}} \, {\mathcal S}\, \ket{S_{MS}}_u 
 +{2\over 18}  \bra{S_{MS}} \, {\mathcal S}\, \ket{S_{MS}}_c  \nonumber \\
&& +{1\over 3}  \bra{S_{MA}} \, {\mathcal S}\, \ket{S_{MA}}_c
\nonumber \\
\bra{\Omega^{+}_{cc} } \; {\mathcal S F} \; \ket{\Omega^{+}_{cc}}&=&
 -{2\over 18}  \bra{S_{MS}} \, {\mathcal S}\, \ket{S_{MS}}_s 
 +{2\over 18}  \bra{S_{MS}} \, {\mathcal S}\, \ket{S_{MS}}_c \nonumber \\
&& +{1\over 3}  \bra{S_{MA}} \, {\mathcal S}\, \ket{S_{MA}}_c \, . 
\eeqa
The non-zero transition matrix elements are:
\beqa
\bra{\Sigma^+_c } \; {\mathcal S F} \; \ket{  \Lambda_c}&=&
  {1\over 2\sqrt{12}}  \bra{S_{MS}} \, {\mathcal S}\, \ket{S_{MS}}_u 
 -{1\over 2\sqrt{12}}  \bra{S_{MA}} \, {\mathcal S}\, \ket{S_{MA}}_u 
\nonumber \\
\bra{\Xi^+_c } \; {\mathcal S F} \; \ket{ \Xi^{+}_{c\bar{3}} }&=&
  {1\over 6\sqrt{3}}  \bra{S_{MS}} \, {\mathcal S}\, \ket{S_{MS}}_u 
  +{1\over 12\sqrt{3}}  \bra{S_{MS}} \, {\mathcal S}\, \ket{S_{MS}}_s\nonumber \\
&&
  -{1\over 6\sqrt{3}}  \bra{S_{MA}} \, {\mathcal S}\, \ket{S_{MA}}_u 
  -{1\over 12\sqrt{3}}  \bra{S_{MA}} \, {\mathcal S}\, \ket{S_{MA}}_s
\nonumber \\
\bra{\Xi^0_c } \; {\mathcal S F} \; \ket{ \Xi^{0}_{c\bar{3}} }&=&
  -{1\over 12\sqrt{3}}  \bra{S_{MS}} \, {\mathcal S}\, \ket{S_{MS}}_u 
  +{1\over 12\sqrt{3}}  \bra{S_{MS}} \, {\mathcal S}\, \ket{S_{MS}}_s \nonumber \\
&&
  +{1\over 12\sqrt{3}}  \bra{S_{MA}} \, {\mathcal S}\, \ket{S_{MA}}_u  
  -{1\over 12\sqrt{3}}  \bra{S_{MA}} \, {\mathcal S}\, \ket{S_{MA}}_s \,.
\eeqa
Here for example, $\bra{S_{MS}} \, {\mathcal S}\,
 \ket{S_{MS}}_c$, stands
for a matrix element between two mixed symmetric states of the 
spin operator ${\mathcal S}$, where the struck quark is a charm quark. 
Depending on the baryon the other two quarks could 
be for example $uu$ ($\Sigma^{++}$) or $us$ ($\Xi^{a+}_c$).

\begin{table}[t]
\caption{$SU(4)_F$ wave functions of the mixed symmetry baryons. We also include
the non-strange ones for completeness\label{bwf4}}
\bce
\begin{tabular}{lcc}
Baryon         &  $\phi_{MS}$                              &   $\phi_{MA}$  \\
\hline
\multicolumn{3}{c}{Mixed symmetry 20plet: (8,0)} \\
\hline
$p$            & ${1\over\sqrt{6}} [uud+udu-2duu]$   &  ${1\over\sqrt{2}}(uud-udu)$\\
$n$            & ${-1\over\sqrt{6}} [ddu+dud-2udd]$   &  ${1\over\sqrt{2}}(dud-ddu)$\\
$\Sigma^+$     & ${1\over\sqrt{6}} [uus+usu-2suu]$   &  ${1\over\sqrt{2}}(uus-usu)$\\
$\Sigma^0$     & ${1\over \sqrt{6}} \left[ { usd+dsu\over\sqrt{2}} 
+  { uds+dus\over\sqrt{2}} - 2{ sdu+sud \over\sqrt{2}} \right]$ &
$ {1\over\sqrt{2}}\left[ { uds+dus \over\sqrt{2}} - { dsu+usd \over\sqrt{2}}   \right]$ \\
$\Sigma^-$     & ${1\over\sqrt{6}} [dds+dsd-2sdd]$   &  ${1\over\sqrt{2}}(dds-dsd)$\\
$\Lambda^0$    &$ {1\over\sqrt{2}}\left[ { uds-dus \over\sqrt{2}} +
 {usd-dsu  \over\sqrt{2}} \right]$ &
${1\over \sqrt{6}} \left[ { usd-dsu\over\sqrt{2}} 
+  { dus-uds\over\sqrt{2}} - 2{ sdu-sud\over\sqrt{2}} \right]$\\
$\Xi^-$        & ${-1\over\sqrt{6}} [sds+ssd-2dss]$   &  ${1\over\sqrt{2}}(sds-ssd)$\\
$\Xi^0$        & ${-1\over\sqrt{6}} [sus+ssu-2uss]$   &  ${1\over\sqrt{2}}(sus-ssu)$\\
\hline
\multicolumn{3}{c}{Mixed symmetry 20plet: (6,1)} \\
\hline
$\Sigma^{++}$     & ${1\over\sqrt{6}} [uuc+ucu-2cuu]$   &  ${1\over\sqrt{2}}[uuc-ucu]$\\
$\Sigma_c^+$     & ${1\over \sqrt{6}} \left[ { ucd+dcu\over\sqrt{2}} 
+  { udc+duc\over\sqrt{2}} - 2{ cdu+cud \over\sqrt{2}} \right]$ &
$ {1\over\sqrt{2}}\left[ { udc+duc \over\sqrt{2}} - { dcu+ucd \over\sqrt{2}}   \right]$ \\
$\Sigma_c^0$     & ${1\over\sqrt{6}} [ddc+dcd-2cdd]$   &  ${1\over\sqrt{2}}[ddc-dcd]$\\
$\Xi_c^{s+}$     & ${1\over \sqrt{6}} \left[ { usc+suc\over\sqrt{2}} 
+  { ucs+scu\over\sqrt{2}} - 2{ cus+csu \over\sqrt{2}} \right]$ &
$ {1\over\sqrt{2}}\left[ { usc+suc \over\sqrt{2}} - { ucs+scu \over\sqrt{2}}   \right]$ \\
$\Xi_c^{s0}$     & ${1\over \sqrt{6}} \left[ { dsc+sdc\over\sqrt{2}} 
+  { dcs+scd\over\sqrt{2}} - 2{ cds+csd \over\sqrt{2}} \right]$ &
${1\over\sqrt{2}}\left[ { dsc+sdc \over\sqrt{2}} - { dcs+scd \over\sqrt{2}}   \right]$ \\
$\Omega^{0}_c$     & ${1\over\sqrt{6}} [ssc+scs-2css]$   &  ${1\over\sqrt{2}}[ssc-scs]$\\
\hline
\multicolumn{3}{c}{Mixed symmetry 20plet: ($\bar{3}$,1)} \\
\hline
$\Lambda_c^+$    &$ {1\over\sqrt{2}}\left[ { udc-duc \over\sqrt{2}} +
 {ucd-dcu  \over\sqrt{2}} \right]$ &
${1\over \sqrt{6}} \left[ { ucd-dcu\over\sqrt{2}} 
+  { duc-udc\over\sqrt{2}} - 2{ cdu-cud\over\sqrt{2}} \right]$\\
$\Xi_c^{a+}$    &$ {1\over\sqrt{2}}\left[ { usc-suc \over\sqrt{2}} +
 {ucs-scu  \over\sqrt{2}} \right]$ &
${1\over \sqrt{6}} \left[ { ucs-scu\over\sqrt{2}} 
+  { suc-usc\over\sqrt{2}} - 2{ csu-cus\over\sqrt{2}} \right]$\\
$\Xi_c^{a0}$    &${1\over\sqrt{2}}\left[ { dsc-sdc \over\sqrt{2}} +
 {dcs-scd  \over\sqrt{2}} \right]$ &
${1\over \sqrt{6}} \left[ { dcs-scd\over\sqrt{2}} 
+  { sdc-dsc\over\sqrt{2}} - 2{ csd-cds\over\sqrt{2}} \right]$\\
\hline
\multicolumn{3}{c}{Mixed symmetry 20plet: (3,2)} \\
\hline
$\Xi^{++}_{cc}$     & ${1\over\sqrt{6}} [ccu+cuc-2ucc]$   &  ${1\over\sqrt{2}}[ccu-cuc]$\\
$\Xi^{+}_{cc}$     & ${1\over\sqrt{6}} [ccd+cdc-2dcc]$   &  ${1\over\sqrt{2}}[ccd-cdc]$\\
$\Omega^{+}_{cc}$     & ${1\over\sqrt{6}} [ccs+csc-2scc]$   &  ${1\over\sqrt{2}}[ccs-csc]$\\
\hline
\end{tabular}
\ece
\end{table}

\end{document}